\documentclass[%
 reprint,
 amsmath,amssymb,
 aps,
]{revtex4-2}

\usepackage{graphicx}
\usepackage{dcolumn}
\usepackage{bm}
\usepackage{bbm}

\usepackage[utf8]{inputenc}  
\usepackage[T1]{fontenc}     

\usepackage{amsmath}
\usepackage{amssymb}
\usepackage{graphicx}
\usepackage{mathrsfs}
\usepackage{amsfonts}
\usepackage{amsthm}
\usepackage{color}
\usepackage{txfonts}
\usepackage[colorlinks=true,citecolor=blue,linkcolor=blue,urlcolor=blue,anchorcolor=blue]{hyperref}%
\usepackage{pifont}
\hypersetup{colorlinks=true,citecolor=blue,linkcolor=blue,urlcolor=blue}
\providecommand{\U}[1]{\protect\rule{.1in}{.1in}}
\makeatletter
\@ifundefined{textcolor}{}
{
\definecolor{BLACK}{gray}{0}
\definecolor{WHITE}{gray}{1}
\definecolor{RED}{rgb}{1,0,0}
\definecolor{GREEN}{rgb}{0,1,0}
\definecolor{BLUE}{rgb}{0,0,1}
\definecolor{CYAN}{cmyk}{1,0,0,0}
\definecolor{MAGENTA}{cmyk}{0,1,0,0}
\definecolor{YELLOW}{cmyk}{0,0,1,0}
}

\begin{document}


\title{Tunable spin-wave nonreciprocity in ferrimagnetic domain-wall channels}

\author{Tingting Liu$^{1}$}
\email[Corresponding author: ]{tingtingliu@lcu.edu.cn}
\author{Shuhong Li$^{1}$}
\author{Yunlong Liu$^{1}$}
\author{Shuchao Qin$^{1}$}

\author{Yang Liu$^{2}$}
\email[Corresponding author: ]{dyly1019@163.com}

\author{Wenjun Wang$^{1}$}
\email[Corresponding author: ]{phywwang@163.com}  

\author{Minghui Qin$^{3}$}

\affiliation{$^{1}$School of Physics Science and Information Technology, Liaocheng University, Liaocheng 252000, China}
\affiliation{$^{2}$School of Physics and State Key Laboratory of Electronic Thin Films and Integrated Devices, University of Electronic Science and Technology of China, Chengdu 610054, China}
\affiliation{$^{3}$Guangdong Provincial Key Laboratory of Quantum Engineering and Quantum Materials and Institute for Advanced Materials, South China Academy of Advanced Optoelectronics, South China Normal University, Guangzhou 510006, China}

\begin{abstract}
The nonreciprocal propagation of spin waves (SWs) offers opportunities for developing novel functional magnonic logic devices, where controllability is crucial for magnetic signal processing. Domain walls act as natural waveguides due to their magnetic configuration, offering a platform for the in-depth investigation of nonreciprocal SW propagation and its manipulation. In this work, we theoretically and numerically investigate the tunable SW nonreciprocity in ferrimagnetic domain-wall channels under the influence of external field. It is revealed that the Dzyaloshinskii-Moriya interaction exerts dual control over both nonreciprocal SW propagation and spin splitting phenomena. Moreover, SW nonreciprocity is magnetically tunable, with its sign reversibly switched by inverting the applied field direction, while preserving the host spin configuration. The orientation of the magnetic field can selectively modify the domain wall structure, offering precise control over SW nonreciprocity. Ultimately, we demonstrate a controllable SW transmission scheme via external magnetic field modulation, providing critical insights for the design of future magnonic devices.

\end{abstract}

\maketitle


\section{\label{sec:level1}INTRODUCTION}

Spin waves (SWs), as collective excitations of magnetic moments in ordered magnetic systems, have garnered significant attention due to their potential applications in information transmission \cite{1PhysRevLett.114.247206,2PhysRevB.103.214420,3PhysRevB.100.024416,4Liu2018}. In contrast to traditional spin-polarized electron-based spintronics, SWs do not require charge transport, thus eliminating Joule heating \cite{5PhysRevLett.112.147204,6PhysRevB.97.020402,7PhysRevB.100.174403,8PhysRevB.104.054419,9PhysRevLett.134.026701}. In principle, SWs can propagate over long distances through magnetic insulators with minimal dissipation, making them promising candidates for energy-efficient information carriers, potentially reducing energy consumption significantly \cite{10Jin_2022,11YU20211,12YUAN20221}.

Remarkably, recent studies show that chiral materials host intrinsically nonreciprocal SW propagation, a symmetry-breaking phenomenon that presents unprecedented opportunities for controlling magnonic information flow \cite{13Han2021,14PhysRevB.88.184404,15PhysRevB.100.104427,16PhysRevLett.124.027203,17An2013,18PhysRevB.103.144411}. For example, nonreciprocal devices could enable the development of nonreciprocal antennas and radomes for full-duplex wireless communication and radar systems \cite{19Yang2023}. Specifically, magnetostatic surface SWs, also known as Damon-Eshbach mode, can generate nonreciprocal SW propagation by magnetic dipole interactions \cite{20DAMON1961308,21PhysRevB.96.014404,DEPhysRevApplied.22.034046}. As the film thickness and SW wavelength decrease, the contribution of magnetic dipole interactions gradually diminishes, leading to the weakening or even disappearance of nonreciprocity. Alternatively, chiral materials exhibit inherent Dzyaloshinskii-Moriya interaction (DMI)-induced nonreciprocal SW propagation when the magnetization and wave vector lie in-plane and are mutually perpendicular, a phenomenon arising from the breaking of inversion symmetry \cite{22PhysRevB.89.224408,23PhysRevB.91.180405,DMPhysRevLett.132.036701}. 

Building on this intrinsic symmetry-breaking mechanism, enhancing the tunability of SW nonreciprocity motivates a promising strategy that leverages chiral ferrimagnets. Near the angular momentum compensation temperature, their reduced net magnetic moment enables rapid dynamics \cite{24PhysRevB.96.100407,25Kim2020} while maintaining detectability through conventional techniques \cite{26PhysRevB.109.174412,27Caretta2018,28PhysRevLett.121.057701,2910.1063/5.0146374}. Crucially, the antiferromagnetic coupling between inequivalent sublattices grants full-polarization degrees of freedom \cite{7PhysRevB.100.174403}, enabling effective SW excitation and manipulation, including nonreciprocal propagation induced by antisymmetric chiral DMI \cite{30PhysRevB.106.224413}. However, the inherent dependence of SW nonreciprocity effect on the chirality of pre-established spin textures poses a significant challenge for developing programmable SW devices. 

\setlength{\parskip}{0pt} 
To address this limitation, a promising strategy is to modify the equilibrium spin configurations by applying an external magnetic field, which can be achieved through integrated current traces or microcoils. While external magnetic fields have been shown to induce SW nonreciprocity in antiferromagnets \cite{31PhysRevB.105.094436}, the underlying theoretical framework remains underdeveloped. Moreover, the net angular momentum inherent to ferrimagnetic systems offers the potential for significantly enhancing information capacity as SWs are used as information carriers. Thus, external magnetic fields provide a viable approach for flexibly tuning SW nonreciprocity in ferrimagnets, with potentially emergent phenomena to be expected. 

\setlength{\parskip}{0pt} 
In this work, we present a theoretical and numerical study of tunable SW nonreciprocity in ferrimagnetic domain-wall channels under an external magnetic field. Both the magnitude and sign of the nonreciprocity are shown to depend sensitively on the field strength, enabling field-controlled modulation of SW transport. The SW mode splitting is further enhanced by the DMI, interlayer exchange coupling, and the net angular momentum $\delta_{s}$. These findings demonstrate that external magnetic fields provide a versatile means to manipulate SW propagation without modifying the chirality of the spin texture, offering a promising route toward reconfigurable magnonic devices.

\section{\label{sec:level2}MODEL AND METHODS}

\begin{figure}[t]
    \centering
    \includegraphics[width=0.45\textwidth]{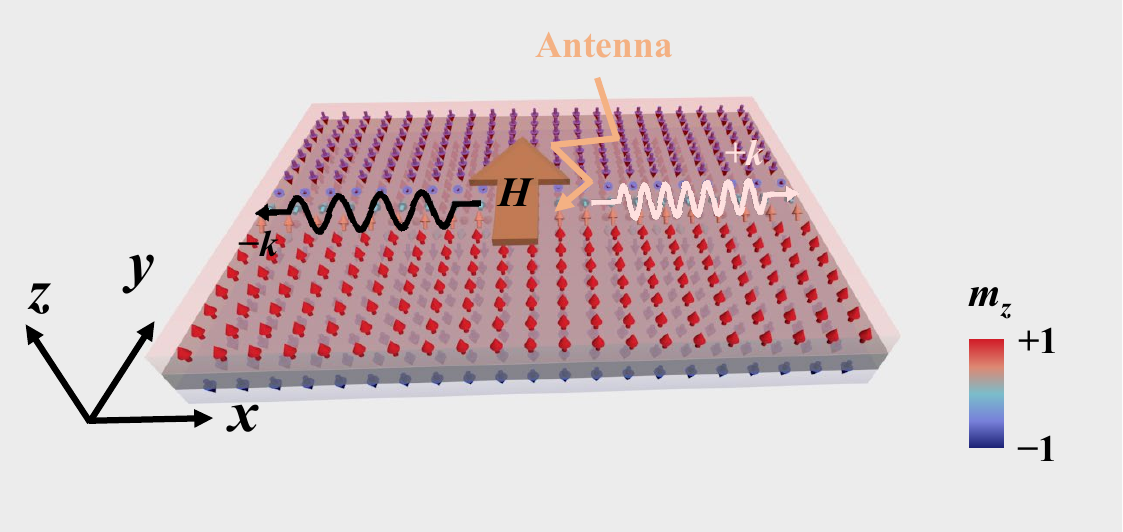}
    \caption{Schematic spin configuration of synthetic ferrimagnetic domain walls. The SW is excited by antenna directed along the $x$ direction. Here, red and blue arrows represent the magnetization direction of the two sublattices.}
    \label{fig1}
\end{figure}
We consider a spin model composed of two unequal sublattices which are antiferromagnetically coupled, as shown in Fig. \ref{fig1}. The unit magnetization vectors of the two sublattices are $\mathbf{m}_1$ and $\mathbf{m}_2$, respectively. In the continuum approximation, one introduces the $\mathrm{N\acute{e}el}$ vector $\mathbf{n}= (\mathbf{m}_1-\mathbf{m}_2)/2$ and the magnetization vector $\mathbf{m}= (\mathbf{m}_1+\mathbf{m}_2)/2$ to deal with the dynamic equations of ferrimagnets. In addition, for the two sublattices, their gyromagnetic ratio and the Gilbert damping constant are respectively denoted as $\gamma_{1,2}$ and $\alpha_{1,2}$. The spin density of sublattice $i(i=1,2)$ is given by $s_i=M_i/\gamma_i$ with $\gamma_i=g_i\mu_B/\hbar$, where $M_\mathrm{i}$ is the saturation magnetization, $g_i$ is the $\mathrm{Land\acute{e}}$ factor, and $\mu_B$ is the Bohr magneton. 

\begin{figure*}[t] 
  \centering
  \includegraphics[width=0.7\textwidth]{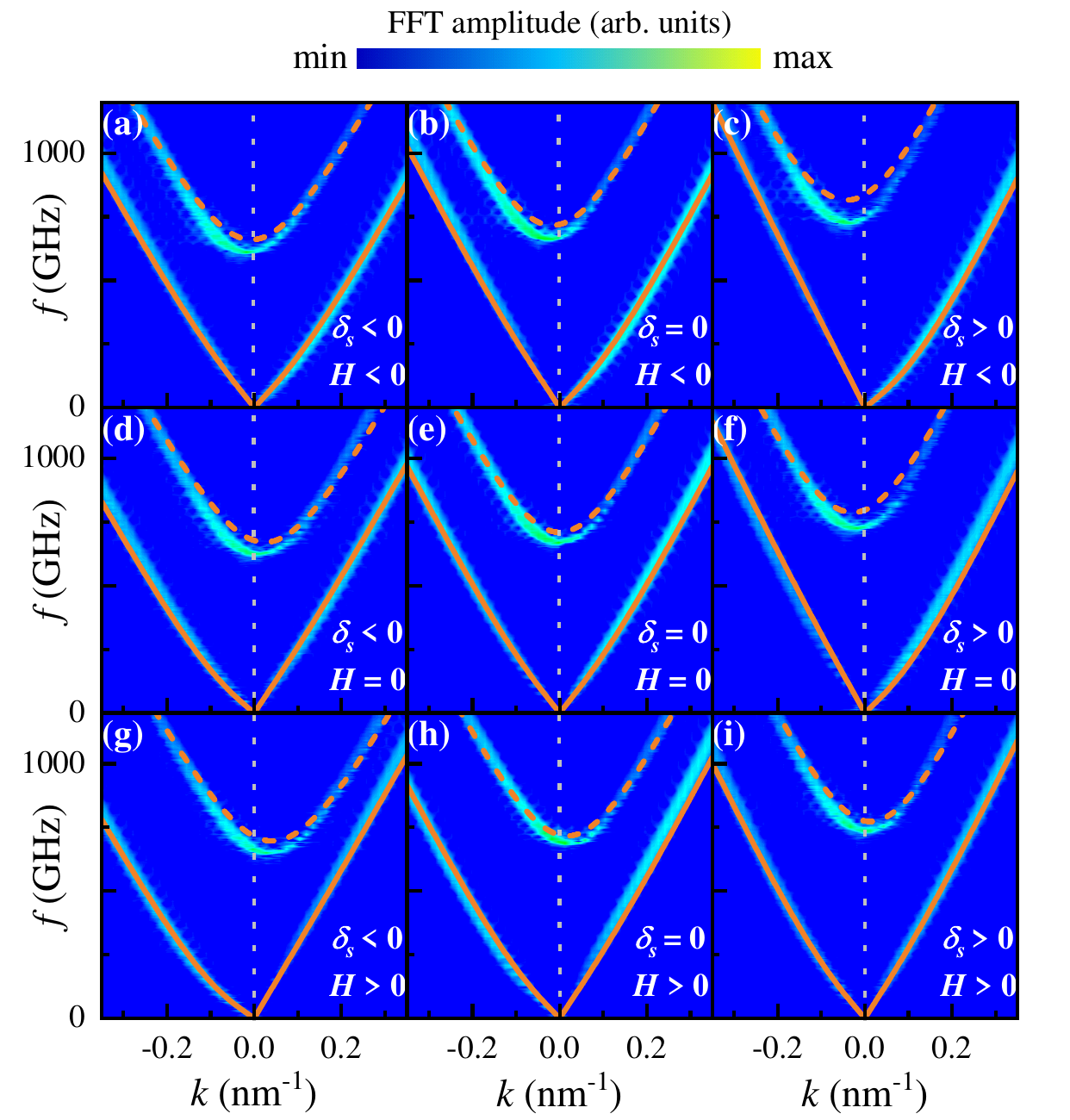} 
  \caption{Ferrimagnetic SW dispersion in domain walls for (a) $\delta_s=-1.24\times10^{-7}\mathrm{~J}\mathrm{s/m}^3,H=-2\mathrm{~T}$, (b) $\delta_s=0,H=-2\mathrm{~T}$, (c) $\delta_s=1.24\times10^{-7}\mathrm{~J}\mathrm{s/m}^3,H=-2\mathrm{~T}$,(d) $\delta_s=-1.24\times10^{-7}\mathrm{~J}\mathrm{s/m}^3,H=0$, (e)$\delta_s=0,H=0$, (f) $\delta_s=1.24\times10^{-7}\mathrm{~J}\mathrm{s/m}^3,H=0$, (g) $\delta_s=-1.24\times10^{-7}\mathrm{~J}\mathrm{s/m}^3,H=2\mathrm{~T}$, (h) $\delta_s=0,H=2\mathrm{~T}$, and (i) $\delta_s=1.24\times10^{-7}\mathrm{~J}\mathrm{s/m}^3,H=2\mathrm{~T}$, with interlayer coupling ${J_{i}=-6\mathrm{~pJ/m}}$. Dashed lines show the analytical results.}
  \label{fig2}
\end{figure*}

Following the standard procedure in earlier work, the Lagrangian density for a ferrimagnetic lattice is given by\cite{32PhysRevB.95.140404}

\begin{equation}\label{eq1}
   L=\frac{\rho(\dot{\mathbf{n}}-g_t\mathbf{n}\times\mathbf{H})}{2}-\delta_s\mathbf{a}(\mathbf{n})\cdot\dot{\mathbf{n}}-U, 
\end{equation} 
where $s=(s_{1}+s_{2})/2$ is the staggered spin density, $\delta_{s}=s_{1}-s_{2}$ is the net angular momentum, $g_{t}$ is the gyromagnetic ratio for the transverse components of the spin density with respect to the magnetization direction, and $\mathbf{a}(\mathbf{n})$ is the vector potential of a magnetic monopole satisfying $\nabla_{\mathbf{n}}\times\mathbf{a}(\mathbf{n})=\mathbf{n}$. The potential energy density $U$ is given \cite{24PhysRevB.96.100407,33PhysRevB.101.024414}
\begin{equation}\label{eq2}
  U=\frac{a}{2}\mathbf{m}^2+\frac{A}{2}\left(\nabla\mathbf{n}\right)^2-\frac{K}{2}n_z^2+\frac{D}{2}[n_z(\nabla\cdot\mathbf{n})-(\mathbf{n}\cdot\nabla)n_z]-M_\mathrm{net}\mathbf{H}\cdot\mathbf{n}, 
\end{equation} 
where $A$ and $\alpha$ are the inhomogeneous and homogeneous exchange constants, respectively, $D$ denotes the interface-induced DMI, $M_\mathrm{net}$ is the net magnetic moment, ${\bf H} = (0, H, 0)$ is the external field applied along the $y$ axis, and $K$ is the effective easy-axis anisotropy. For simplicity, the nonlocal dipolar interaction incorporated into the anisotropy $K$, considering the small net magnetization in the investigated systems. The Rayleigh dissipation function is given by $R=s_\alpha\dot{\mathbf{n}}^2/2$, where $s_a=\alpha_1s_1+\alpha_2s_2$ is a phenomenological parameter quantifying the energy and spin loss due to the magnetic dynamics, and we assume the Gilbert damping constant $\alpha_1=\alpha_2=\alpha$ in this work. The dynamic variable  ${\bf m}$ can be expressed by $\mathbf{m}=-(s/a)\mathbf{\dot{n}}\times\mathbf{n}$. From the Lagrangian density and the Rayleigh dissipation, we obtain the equation of motion in terms of staggered vector ${\bf n}$ by integrating out the net magnetization variable ${\bf m}$
\begin{equation}\label{eq3}
  \rho\mathbf{n}\times\mathbf{\ddot{n}}+\delta_s\dot{\mathbf{n}}+2\alpha s\mathbf{n}\times\dot{\mathbf{n}}-2\rho\mathbf{g}_t\dot{\mathbf{n}}\left(\mathbf{H}\cdot\mathbf{n}\right)=\mathbf{n}\times\mathbf{f}_\mathbf{n}, 
\end{equation} 
here, $\rho = s^2/ a$ parametrizes the inertia, and $\mathbf{f_n}=-\delta U/\delta\mathbf{n}$ is the effective field. We next look into the propagation of SWs in a ferrimagnetic domain wall. We ignore the damping term and consider a small fluctuation of the staggered vector $\mathbf{n}$ around the initial $\mathbf{n}_0$, $\mathbf{n}=\mathbf{n}_0+\delta\mathbf{n}e^{i(\omega t-kx)}$, where $\omega$  and $k$ are the SW frequency and wave vector, respectively. We consider the Walker ansatz for a $\mathrm{N\acute{e}el}$-type DW profile centered at $y=0$, i.e., $\mathbf{n}_0=[0,-$sin$\theta$, cos$\theta]$ with  $\theta=2\arctan[\exp(y/\Delta)]$, where $\Delta=(A/K)^{1/2}$ is the domain wall width. Subsequently, we choose to transform the coordinates with $\mathbf{i}=[1,0,0]$ and $\mathbf{j}=\mathbf{n}_0\times\mathbf{i}$  for the sake of calculation simplicity. By arranging Eq. \eqref{eq3} based on i and j, one obtains
\begin{equation}\label{eq4a}
  -\rho\delta\ddot{n}_j-(\delta_s-\rho\xi)\delta\dot{n}_i=\mathcal{D}_1\delta n_i+\mathcal{U}\delta n_j,\tag{4a} 
\end{equation} 
\begin{equation}\label{eq4b}
  \rho\delta\ddot{n}_i-(\delta_s-\rho\xi)\delta\dot{n}_j=\mathcal{D}_1\delta n_j-(\mathcal{U}+\mathcal{D}_2)\delta n_i,  \tag{4b}
\end{equation} 
Here, $\xi=2g_{t}H\sin\theta$, $\mathcal{U}=-A\nabla^{2}+K(1-2\sin^{2}\theta)-M_{\text{net}}H\sin\theta$, $\mathcal{D}_{1}=D\sin\theta\partial_{x}$, and $\mathcal{D}_{2}=D\sin\theta/\Delta$. We ignore the Zeeman term induce by the weak net magnetization $M_{\text{net}}$, and solve Eqs. \eqref{eq4a} and \eqref{eq4b} by using the perturbation theory and regarding the Winter mode as a scattering basis. Eqs. \eqref{eq4a} and \eqref{eq4b} can be updated to
\begin{equation}\label{eq5a}
  \left(V-\rho\omega^2\right)\delta n_j-\left[\left(\delta_s-\rho\zeta\right)i\omega+iR_1\right]\delta n_i=0,\tag{5a} 
\end{equation} 
\begin{equation}\label{eq5b}
  \left(\rho\omega^2-V-R_2\right)\delta n_i-\left[\left(\delta_s-\rho\zeta\right)i\omega+iR_1\right]\delta n_j=0,  \tag{5b}
\end{equation} 
where $V=Ak^{2}$, $\zeta=2g_{i}H$, $R_{1}=\pi Dk/4$, and $R_{2}=\pi D/4\Delta$. By defining a complex field as $\psi_{\pm}=\delta n_{i}\mp\delta n_{j}$ for right-handed (+) and left-handed ($-$) SWs, we obtain 
\begin{equation}\label{eq6}
  \left(\delta_s-\rho\zeta\right)\omega_\pm-R_1\mp q\left(-\rho\omega_\pm^2+V+R_2\right)=0,  \tag{6}
\end{equation} 
where $q=[(V-\rho\omega_\pm^2)/(V+R_2-\rho\omega_\pm^2)]^{1/2}$. Meanwhile, we can get the nontrival solution of the magnonic system according to Eq. \eqref{eq5a} and \eqref{eq5b},
\begin{multline}\label{eq7}
    -\rho^2\omega^4+\left(2V+R_2+\rho\zeta^2-2\delta_s\zeta+\frac{\delta_s^2}{\rho}\right)\rho\omega^2 \\
  +\left(\delta_s-\rho\zeta\right)2R_1\omega+\left(R_1^2-V^2-R_2V\right)=0. \tag{7}
\end{multline}

Eq. \eqref{eq7} displays the characteristic features of the field-induced nonreciprocity of SWs in ferrimagnetic domain-wall channels. Clearly, the dispersion relations of SWs are closely related to the external magnetic field $H$, the DMI, the inertia coefficient $\rho$, and the net angular momentum $\delta_{s}$ of ferrimagnets. The presence of nonreciprocity is manifested by the third term, i.e., $(\delta_{s}-\rho\zeta)2R_{1}\omega$, in Eq. \eqref{eq7}, as the other terms exhibit a dependence on even powers of the wave vector $k$. Notably, the DMI, $H$ and $\delta_{s}$ plays an essential role in enabling nonreciprocal propagation of SWs. Crucially, the presence of DMI is indispensable, while the opposing signs of the external magnetic field $H$ and net angular momentum $\delta_{s}$ lead to enhanced nonreciprocal effects. In particular, the orientation of the magnetic field can selectively modify the domain wall structure, offering precise control over SW nonreciprocity. For $D<0$, the third term is rewritten as  $-(\delta_{s}-\rho\zeta)2R_1\omega$  in Eq. \eqref{eq7} due to the initial spin configuration $\mathbf{n}_0=[0,\mathrm{sin}\theta,\mathrm{cos}\theta]$. Thus, the sign of the DMI does not affect the main conclusions of this work, and we consider $D>0$ if not stated otherwise. Crucially, while both $R_{1}$ and $R_2$ originate from DMI, their functional manifestations exhibit significant distinctions, as analyzed in the following section.
\begin{figure*}[t] 
  \centering
  \includegraphics[width=0.8\textwidth]{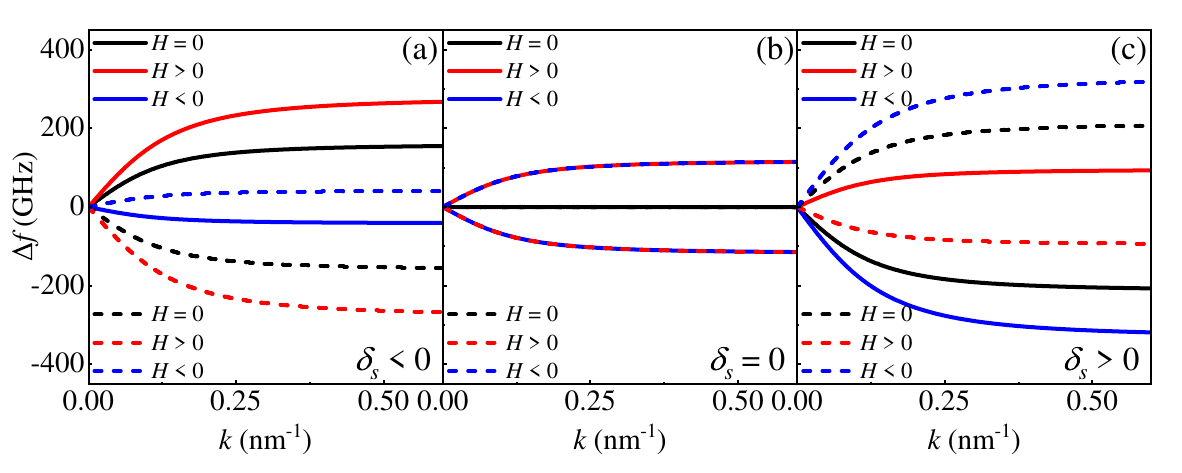} 
  \caption{The frequency shift $\Delta f$ as function of wave vector $k$ under different external magnetic field at (a) $\delta_{s}<0$, (b) $\delta_{s}=0$, and (c) $\delta_{s}>0$ with $J_{i}=-6\mathrm{~pJ/m}$, $D=2.5 \mathrm{~mJ/m^2}$.The solid and dashed lines correspond to the acoustic mode and optical mode, respectively.}
  \label{fig3}
\end{figure*}

\setlength{\parskip}{0pt} 
Furthermore, we perform numerical simulations of the atomistic spin model based on the atomistic Landau-Lifshitz-Gilbert (LLG) equation, in order to check the validity of the theoretical treatment. We use the saturation magnetization listed in Table \ref{tab:table1}, and the simulation details and parameter choice are presented in the Appendix.

\section{\label{sec:level3}RESULTS AND DISCUSSION}

We first discuss the nonreciprocity of SW in absence of external magnetic field. Notably, the dispersion curves of SWs  are symmetric about the $k$ axis at the angular momentum compensation point $\delta_{s}=0$, as shown in Fig. \ref{fig2}(e). This symmetry demonstrates that the SW propagation retains reciprocity within ferrimagnetic domain walls, even in the presence of DMI. For $\delta_{s}\neq 0$, a pronounced nonreciprocity appears even when $H=0$, as illustrated in Figs. \ref{fig2}(d) and (f). This nonreciprocity is reflected in the third term of Eq. \eqref{eq7}, ie. $(\delta_{s}-\rho\zeta)2R_1\omega$, as the other terms are proportional to the even power of the wave vectors $k$. 

\setlength{\parskip}{0pt} 
We next discuss the tunable SW nonreciprocity in ferrimagnetic domain walls channels by external field $H$. A pronounced nonreciprocal effect emerges under finite magnetic fields $(H\neq0)$ at $\delta_{s}=0$, contrasting with the approximately symmetric dispersion relations at zero field $(H=0)$, as shown in Figs. \ref{fig2}(b)(e), and (h). This phenomenon arises from the breaking of spatiotemporal inversion symmetry by the combined effects of the magnetic field and DMI. 
This aligns with the theoretical prediction from the third term ($\delta_{s}-\rho\zeta$)2$R_{1}\omega$ in Eq. \eqref{eq7}, where the nonreciprocal term remains finite under coexisting DMI and magnetic fields even when $\delta_{s}=0$. Notably, the DMI-driven nonreciprocity critically depends on the relative sign alignment between the net angular momentum $\delta_{s}$ and the external magnetic field  $H$. Oppositely signed $\delta_{s}$ and $H$ induce enhanced SW nonreciprocity, as demonstrated in Figs. \ref{fig2}(c) and (g). And simultaneous sign reversal of both the magnetic field $H$ and parameter $D$ induces inversion of the nonreciprocal response, as theoretically expected. 
\begin{table}[b]
\vspace{-1.5ex}
\caption{\label{tab:table1}%
Parameters used in the numerical simulation.}
\begin{ruledtabular}
\begin{tabular}{lccc}%
\textrm{Index}&
\textrm{1}&
\textrm{2}&
\textrm{3}\\
\colrule
$M_1~(\mathrm{kA/m})$ & 460 & 440 & 420\\
$M_2~(\mathrm{kA/m})$ & 440 & 400 & 360\\
$\delta_s~(\times10^{-7}~\mathrm{J}\mathrm{s}/\mathrm{m}^3)$ & -1.24 & 0 & 1.24\\
\end{tabular}
\end{ruledtabular}
\end{table}

\setlength{\parskip}{0pt} 
To intuitively demonstrate the nonreciprocity of the SW, we describe the nonreciprocity of the frequency by the frequency difference of the channeled SWs with opposite wave vectors $\Delta f=f(+k)-f(-k)$. The frequency shift $\Delta f$ is presented as function of wave vector $k$ with various parameters in Fig. \ref{fig3}, and the solid and dashed lines correspond to the acoustic mode and optical mode, respectively.
The black curve of the Fig. \ref{fig3}(b) indicates that the DMI almost does not cause frequency nonreciprocity of magnon at the angular momentum compensation point $\delta_{s}=0$ with zero magnetic field $(H=0)$ for both optical and acoustic modes. However, the nonreciprocity appears with $\delta_{s}\neq0$ or $H\neq0$. The observed nonreciprocity arises from the synergistic effect of $\delta_{s}$, $H$ and DMI parameters, and the nonreciprocity is maximized when the signs of $\delta_{s}$ and $H$ are opposite, as show in Figs. \ref{fig3}(a) and \ref{fig3}(c). This conclusion corresponds to the third term of Eq. \eqref{eq7}, $(\delta_{s}-\rho\zeta)2R_{1}\omega$. 
Moreover, it can be observed that the nonreciprocal effect manifests in opposite directions within the acoustic and optical modes, as indicated by the solid and dashed lines in Fig. \ref{fig3}. This behavior originates from the opposite signs of $H$ and $\omega$ among the three terms in Eq. \eqref{eq7}.
Thus, the DMI is indispensable for the nonreciprocal effect in ferrimagnetic domain wall channel, and the synergistic coupling between $\delta_{s}$ and magnetic fields $H$ enables flexible tuning nonreciprocity.
 Consequently, the SW nonreciprocity in ferrimagnetic domain-wall channel can be tuned via an external magnetic field for fixed DMI and $\delta_s$, while the inherent constraints of static structures pose challenges for active modulation of nonreciprocity. 
Thus, external magnetic fields enable active modulation of SW nonreciprocity in ferrimagnetic domain-wall channels under fixed DMI and $\delta_s$,, overcoming the chiral constraints intrinsic to spin textures.
 Compared to the DMI, a simpler and more feasible method involves altering the equilibrium spin states through the external magnetic field, which can be generated via antenna or microcoils. Therefore, external magnetic fields provide a viable approach for flexibly tuning SW nonreciprocity, and related SW logic devices can be designed.
\begin{figure*}[t] 
  \centering
  \includegraphics[width=0.7\textwidth]{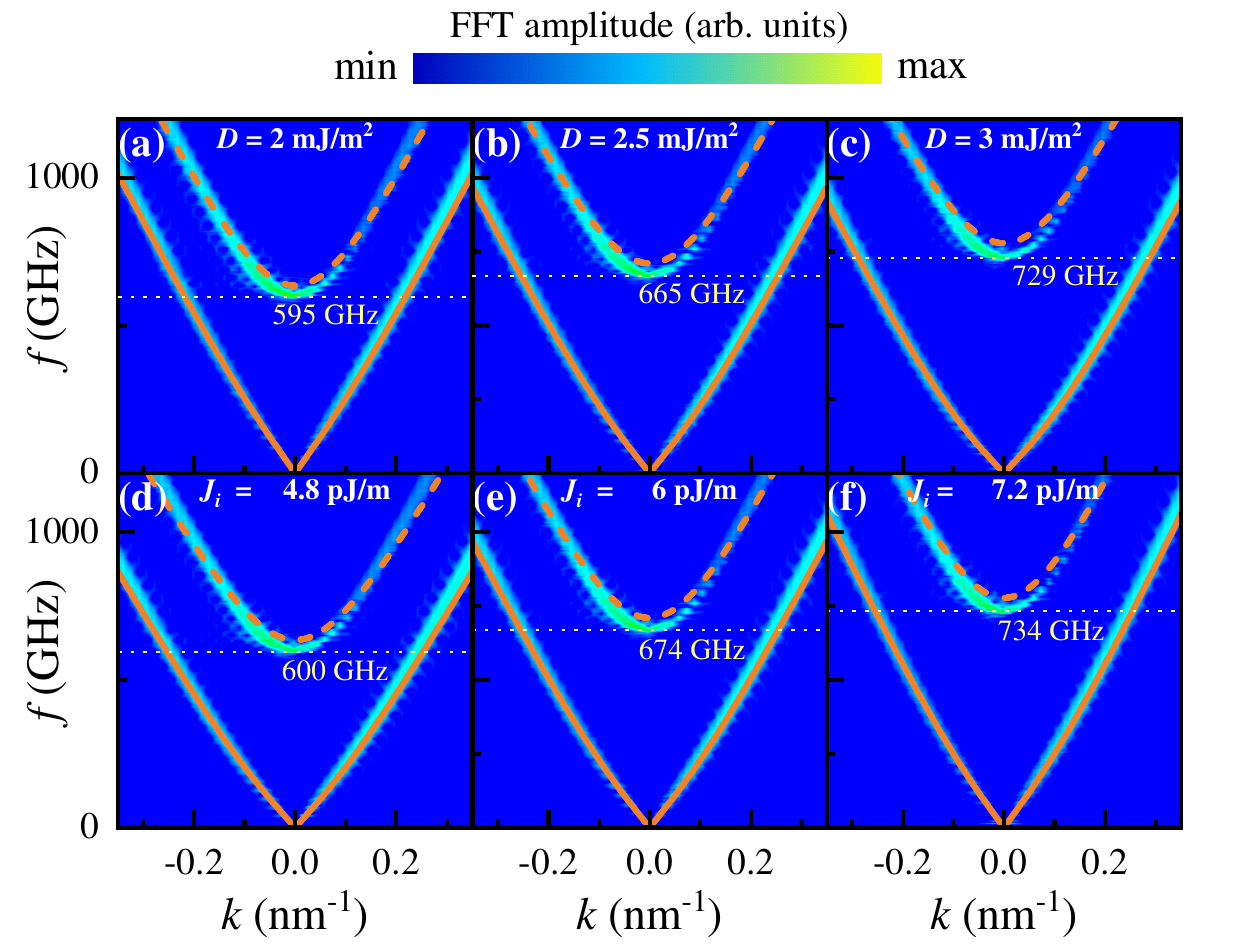} 
  \caption{Ferrimagnetic SW dispersion in domain walls for (a) $D=2 \mathrm{~mJ/m^2}$,(b)\ $D=2.5 \mathrm{~mJ/m^2}$, and (c) $D=3 \mathrm{~mJ/m^2}$ with interlayer coupling $J_{i}=-6\mathrm{~pJ/m}$ and $\delta_{s} = 0$ . (d) $J_{i}=-4.8\mathrm{~pJ/m}$, (e) $J_{i}=-7.2\mathrm{~pJ/m}$ with DMI coefficient $D=2.5 \mathrm{~mJ/m^2}$ and $\delta_{s} = 0$. Dashed lines show the analytical results.}
  \label{fig4}
\end{figure*}

\setlength{\parskip}{0pt} 
Moreover, the spin splitting effect can be observed by SW dispersion relations at the angular momentum compensation point $\delta_{s}=0$, as shown in Fig. \ref{fig2}(e). The spin splitting originates from SW chirality non-degeneracy, characterized by the SW has an acoustic mode without a gap and another optical mode with an energy gap. This different from our previous research results \cite{34PhysRevB.105.214432}, which attributed to the influence of the DMI. The DMI introduces spin splitting by lifting spin degeneracy, which breaks spin symmetry and shifts SW bands for different spin polarizations.The effect of the DMI on spin splitting is investigated by analyzing the DMI coefficient dependent SW dispersion.
For convenience, we use the frequency gap $\omega_{\mathrm{gap}}$ to describe the energy gap between the acoustic and optical modes. Based on Eq. \eqref{eq7}, the frequency gaps $\omega_{\mathrm{gap}}$ can be estimated with the wave vector $k=0$
\begin{equation}\label{eq8}
  \omega_{\mathrm{gap}}=\sqrt{\frac{\pi D}{4\Delta\rho}+4g_t^2H^2-\frac{4\delta_sg_tH}{\rho}+\frac{\delta_s^2}{\rho^2}}.  \tag{8}
\end{equation} 

Eq. \eqref{eq8} shows that the frequency gaps $\omega_{\mathrm{gap}}$ depends on the DMI, the inertia coefficient  $\rho$, and net angular momentum $\delta_{s}$. The external magnetic field $H$ exhibits negligible effects due to the intrinsically small gyromagnetic ratio $g_t$. From the first term of Eq. (\ref{eq8}) (derived from  $R_2$ in Eq. \eqref{eq7}), it can be seen that the frequency gap is proportional to the DMI and increases with the increase of  $D$. Figs. \ref{fig4}(a-c) present the simulated and calculated SW dispersion with $H=0$ for various DMI coefficient in a compensated ferrimagnet $\delta_{s}=0$, which exhibits good agreement between simulations and theory. With the enhancement of the DMI, the frequency gap is significantly widened. This can be understood as the gap between the acoustic and optical branches being opened by the DMI, and the energy gap becomes more pronounced at higher DMI strengths. This is due to the DMI breaking spatial symmetry.
Specifically, the DMI exerts a dual role in modifying the dispersion of SWs: (i) induces nonreciprocal SW dispersion through symmetry breaking, which create asymmetric propagation for $+k$ and $-k$ waves by breaking spatial inversion symmetry, and (ii) introduces spin splitting by lifting spin degeneracy, which breaks spin symmetry and shifts magnon bands for different spin polarizations. In this regard, the nonreciprocity and spin splitting  of the SW are primarily governed by $R_1$ and $R_2$ in Eq. \eqref{eq7}, respectively.

According to  Eq. \eqref{eq8}, the frequency gap is also closely related to $\delta_{s}$ and $\rho$. Finite $\delta_{s}$ breaks the degeneracy of SW modes and concurrently opens a frequency gap in ferrimagnetic domain walls, as established in our previous work \cite{34PhysRevB.105.214432}. Under the condition of vanishing magnetic field and DMI $(H=0,D=0)$ as specified in Eq. \eqref{eq8}, the frequency gap can be described by $\omega_\mathrm{gap}=\delta_s/\rho$, which align with these established conclusions. The functional dependence of the frequency gap on $\delta_{s}$, $\omega_\mathrm{gap}\left(\delta_{s}>0\right)>\omega_\mathrm{gap}\left(\delta_{s}=\right.$ $0)>\omega_{\mathrm{gap}}\left(\delta_{s}<0\right)$, can be predicted by Eq. \eqref{eq8}, which is consistent with the results in Figs. \ref{fig2}(d-f).
Moreover, the inverse proportionality between the inertia coefficient $\rho$ and interlayer coupling coefficient $J_{i}$ implies that increasing $J_{i}$ enlarges the frequency gap  $\omega_{\mathrm{gap}}$. As shown in Fig. \ref{fig4}(d-f), the dispersion curves under varying interlayer coupling strengths reveal a monotonic enlargement of the frequency gap  $\omega_{\mathrm{gap}}$ with increasing coupling coefficient $J_{i}$, in quantitative agreement with theoretical predictions. This phenomenon can be attributed to the fact that the enhanced interlayer exchange field suppresses the excitation of low-energy magnons, similar to the case of antiferromagnets.

Significantly, the precise control of the spacer layer thickness enables modulation of interlayer coupling strength \cite{35https://doi.org/10.1002/adfm.202107490}, the DMI can be modulated through tuning the atomic-scale modulation of interfaces or locally varying the layer thickness \cite{36Samardak2020,37Torrejon2014,38Chen2013,39PhysRevLett.118.147201}. Moreover, as one of the most important parameters in ferrimagnets, $\delta_{s}$ can be elaborately adjusted by tuning the temperature or material composition in experiments \cite{40Kim2022,41PhysRevLett.122.057204,42Hirata2019}. Thus, the frequency gaps between the acoustic and optical modes can be enhanced as the interlayer coupling, DMI, or $\delta_{s}$ increases, which could be used to alleviate the possible interference of unwanted low-frequency magnon modes.
\begin{figure}[t]
    \centering
    \includegraphics[width=0.5\textwidth]{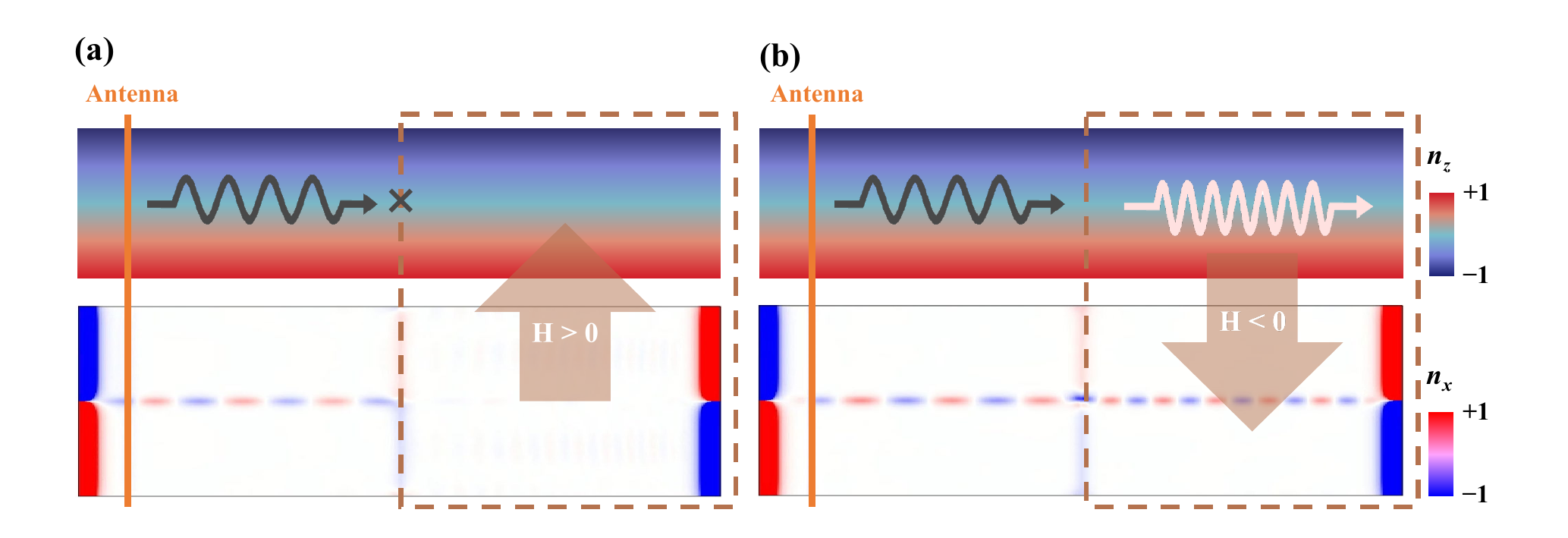}
    \caption{The tuned propagation of SW by the external magnetic field of (a) $H>0$ and (b) $H<0$. Here, $\delta_{s}=0$, $J_{i}=-6\mathrm{~pJ/m}$, and $D=2.5 \mathrm{~mJ/m^2}$.}
    \label{fig5}
\end{figure}

So far, this study unveils the important role of the external magnetic field in tuning the nonreciprocity of SW in the ferrimagnetic domain walls, which is helpful in guiding future experiments and SW device design. Generally, external magnetic fields can be readily adjusted via voltage or electric field application, whereas altering the DMI is highly challenging because of the fixed chirality inherent in the spin texture.
Finally, it is worth pointing out that a precise manipulation of SW nonreciprocity is important for implementing SW logic devices \cite{15PhysRevB.100.104427,43Jamali2013}, such as magnetron transistor. For the application of SW logic devices, we show here a controllable signal transmission scheme, as depicted in Fig. \ref{fig5}. Here, an external magnetic field is applied to tune the signal transmission of the SW. The SW signal will be truncated with the application of a positive magnetic field ($H>0$), as shown in Fig. \ref{fig5}(a). This "truncation region" of the magnetic field control enables the switching on and off of the SW signal.
Field reversal shifts SW signals to shorter wavelengths (Fig. \ref{fig5}(b)), extending the usable bandwidth and thereby increasing achievable channel capacity.

\section{\label{sec:level4}CONCLUSION}
In conclusion, we have theoretically and numerically investigated SW propagation in ferrimagnetic domain-wall channels. The nonreciprocal features of the SW spectrum are shown to be tunable via an external magnetic field, even with fixed material parameters such as the DMI and net angular momentum $\delta_{s}$. We clarify the role of DMI in governing nonreciprocity and spin splitting, revealing its dual influence on both spectral asymmetry and mode structure. The spin splitting arises from the combined effects of DMI, interlayer exchange coupling, and $\delta_{s}$, leading to gapless acoustic and gapped optical modes. These results demonstrate that external magnetic fields offer a flexible and effective means to modulate nonreciprocal SW dynamics in ferrimagnetic systems, with direct implications for magnonic device engineering and experimental realization.

\begin{acknowledgments}
The work is supported by the Natural Science Foundation of China (Grants No. U22A20117, No. 52371243, and No. 61775089), the China Postdoctoral Science Foundation (Grant No. 2025M773378), the Youth Innovation Team for Universities of Shandong Province (Grant No. 2023KJ209), the Natural Science Foundation of the Shandong Province (Grants No. ZR2020KB018, and No. ZR2022MF240), and the Liaocheng University Start-up Fund for Doctoral Scientific Research (Grant No. 318052379).
\end{acknowledgments}

\appendix

\section{Micromagnetic Simulations}

In order to check the validity of the theory, we also perform the numerical simulations of the discrete model. The micromagnetic simulations are performed based on the classical Heisenberg model, and the model Hamiltonian is given by
\begin{equation}
\begin{aligned}
H&=A\left(\nabla\mathbf{m}_i\right)^2+K\left[1-\left(\mathbf{m}_i\cdot\hat{\mathbf{e}}_z\right)^2\right]+D\left[m_i^z\nabla\cdot\mathbf{m}_i-\left(\mathbf{m}_i\cdot\nabla\right)m_i^z\right]\\
&+J_i\mathbf{m}_i\cdot\mathbf{m}_j+g_i\mu_B\mu_0\sum\mathbf{H}\cdot\mathbf{m}_i+H_{demag},
\end{aligned}
\end{equation}
where $\mathbf{m}_i$ represents the local unit magnetization vector with the layer index $i$. $A$, $D$, and $K$ are the intralayer ferromagnetic exchange, DMI, and perpendicular magnetic anisotropy constants, respectively. $J_i$ is the antiferromagnetic interlayer coupling, $H_{demag}$ is the demagnetization energy, and $\mathbf{H}$ is the external field applied along the $y$ axis.

Then, the propagation of SW is investigated by solving the LLG equation,
\begin{equation}\frac{\partial\mathbf{m}_i}{\partial t}=-\gamma_i\mathbf{m}_i\times\mathbf{H}_{\mathrm{eff,}i}+\alpha_i\mathbf{m}_i\times\frac{\partial\mathbf{m}_i}{\partial t},
\end{equation}
where $\mathbf{H}_{\mathrm{eff},i}=M_{i}^{-1}\partial H/\partial\mathbf{m}_{i}$ is the effective field with the magnetic moment $M_{i}$ at site $i$, and the gyromagnetic ratio $\gamma_{i}=g_{i}\mu_{B}/\hbar$ with the g-factors $g_{1}=2.2$ and $g_{2}=2$.

The coupling parameters are chosen to be the same as those in the theoretical analysis. Herein, we consider the material parameters of a GdCo/Co bilayer, where the coupling between the two magnetic layers can be modulated by tuning the thickness of the spacer \cite{26PhysRevB.109.174412}. Without loss of generality, we set the $A=15\mathrm{~pJ/m}$, the DMI constant $D = 2.5 \mathrm{~mJ/m}$, the perpendicular anisotropy parameter $K=1 \mathrm{~MJ/m^{3}}$, and the Gilbert damping coefficient $\alpha_1 = \alpha_2 = 0.001$, unless other wise specifed.

Here, the micromagnetic simulations are performed using the MUMAX3 \cite{mumax10.1063/1.4899186}. The simulations are performed on a ferrimagnetic bilayer GdCo/Co with a size of $512\mathrm{~nm}\times256\mathrm{~nm}\times2\mathrm{~nm}$ and cell size of $1\mathrm{~nm}\times1\mathrm{~nm}\times1\mathrm{~nm}$, and set the time step to $5\times10^{-15}$s, noting that GdCo and Co are widely used materials.

\nocite{*}
\bibliography{apssamp}

\end{document}